\documentclass[10pt,twocolumn]{article} 
\usepackage{simpleConference}
\usepackage{times}
\usepackage{graphicx}
\usepackage{amssymb}
\usepackage{pdflscape}
\usepackage{subcaption}

\begin{document}

\title{Classification of dry age-related macular degeneration and diabetic macular edema from optical coherence tomography images using dictionary learning}

\author{{E. Mousavi} ,
	{R. Kafieh$^{*}$} ,
	{H. Rabbani}
\\
School of Advanced Technologies in Medicine, \\ Isfahan University of Medical Sciences,\\ Isfahan, Iran
\\
{r\_kafieh@yahoo.com}
}
\maketitle
\thispagestyle{empty}

\begin{abstract}
Age-related Macular Degeneration (AMD) and Diabetic Macular Edema (DME) are the major causes of vision loss in developed countries. Alteration of retinal layer structure and appearance of exudate are the most significant signs of these diseases. With the aim of automatic classification of DME, AMD and normal subjects from Optical Coherence Tomography (OCT) images, we proposed a classification algorithm. The two important issues intended in this approach are, not utilizing retinal layer segmentation which by itself is a challenging task and attempting to identify diseases in their early stages, where the signs of diseases appear in a small fraction of B-Scans. We used a histogram of oriented gradients (HOG) feature descriptor to well characterize the distribution of local intensity gradients and edge directions. In order to capture the structure of extracted features, we employed different dictionary learning-based classifiers. Our dataset consists of 45 subjects: 15 patients with AMD, 15 patients with DME and 15 normal subjects. The proposed classifier leads to an accuracy of 95.13\%, 100.00\%, and 100.00\% for DME, AMD, and normal OCT images, respectively, only by considering the 4\% of all B-Scans of a volume which outperforms the state of the art methods. 
\end{abstract}

\section{Introduction }\label{sec1}

\begin{figure}[!t]
	\centering
	\begin{subfigure}[b]{0.48\linewidth}
		\includegraphics[width=\linewidth]{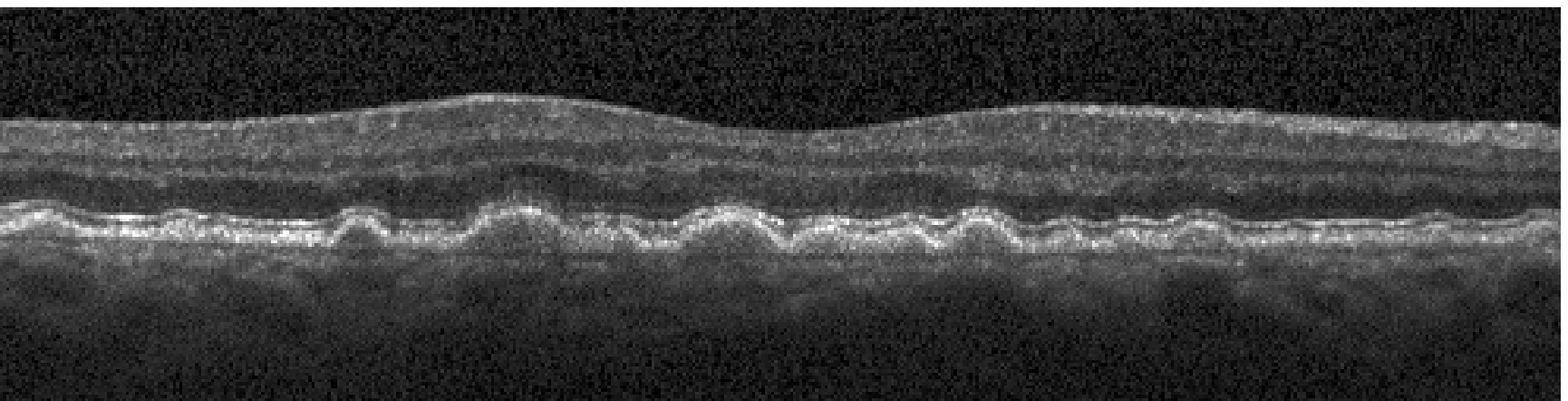}
		\caption{}
		\label{fig_1:1}
	\end{subfigure}
	~ 
	\begin{subfigure}[b]{0.48\linewidth}
		\includegraphics[width=\linewidth]{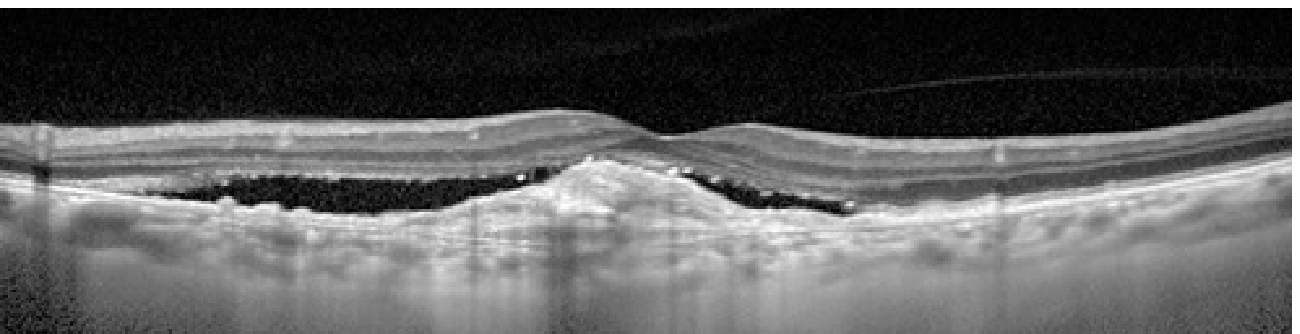}
		\caption{}
		\label{fig_1:2}
	\end{subfigure}
	~ 
	\begin{subfigure}[b]{0.48\linewidth}
		\includegraphics[width=\linewidth]{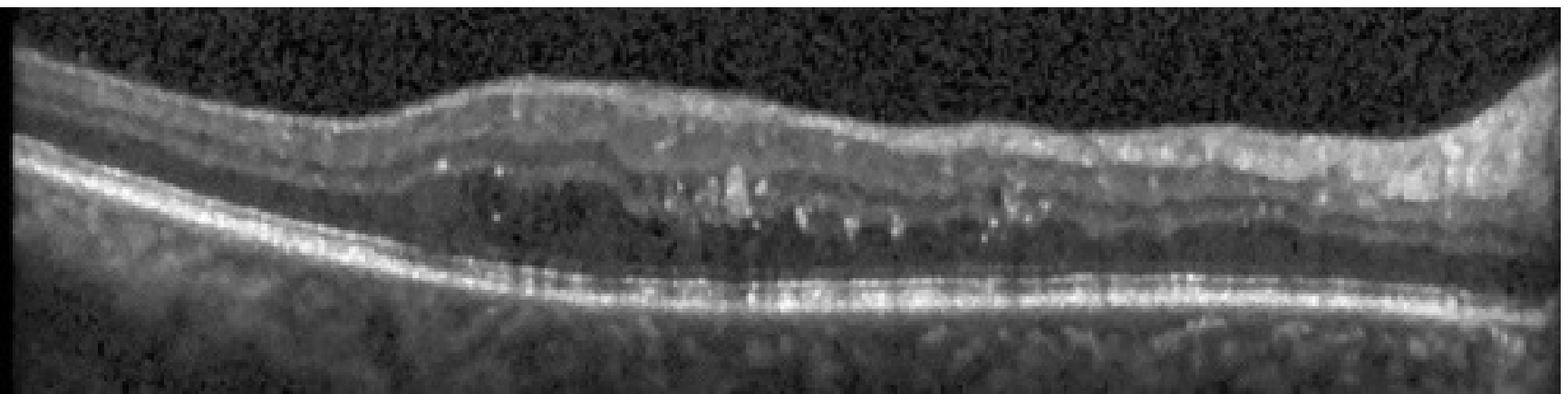}
		\caption{}
		\label{fig_1:3}
	\end{subfigure}
	\begin{subfigure}[b]{0.48\linewidth}
		\includegraphics[width=\linewidth]{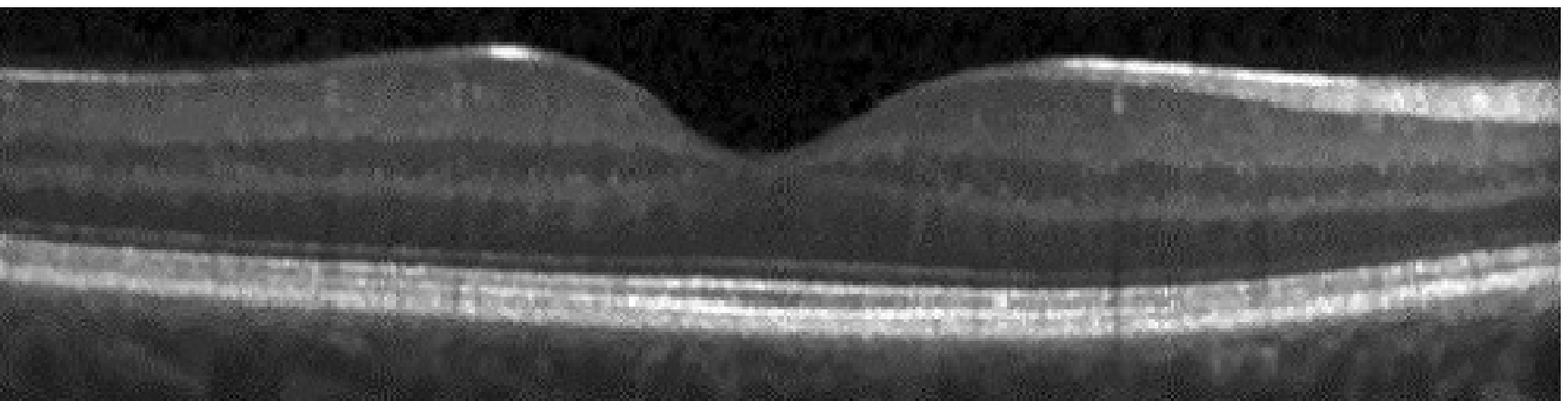}
		\caption{}
		\label{fig_1:4}
	\end{subfigure}
	\caption{Examples of SD-OCT images: a) dry AMD, b) wet AMD, c) DME and d) normal B-scans. }\label{fig_1}
\end{figure}

One of the persistent diseases of the retina is Age-related Macular Degeneration (AMD) which can result in loss of central vision. AMD affects the macula which is an area consist of light-sensitive cells and is responsible for providing sharp central vision. It is estimated that more than 250000 adults suffer blindness due to AMD in United Kingdom \cite{1}. AMD is the term applied to changes in the eye which can be identified by pigmentary abnormalities and considerable drusen, in people aged over 50 years. Drusens are accumulations of lipid material below the retinal pigment epithelium (RPE) and within the Bruch’s membrane, appearing as yellow spots on the retina. AMD disease categorize into two main types: dry form and wet form. Dry form which involves 90\% of people with AMD, is a condition in that layers of the macula (including the RPE and the photoreceptors) get progressively thinner. This is called atrophy. The symptoms of the early stage of dry AMD is a change in the pigment or color of the macula plus appearance of tiny drusen on the retina. Existence of Drusen can lead to deterioration and atrophy of the retina. Exuding is the leakage of fluids from blood vessels and when dry AMD does not involve them, it is also called non-exudative AMD. In contrast to dry AMD, there is another kind of AMD where the growth of new weak blood vessels behind the retina can cause serious problems. This situation which is called wet AMD is accompanied by leakage of fluid, lipids and blood from newly grown vessels which can cause scar tissue to form and retinal cells to stop functioning (see Fig. \ref{fig_1}). Wet AMD is also exudative AMD due to involving exudation or fluid and blood leakage from new blood vessels \cite{2}. 

Diabetic Macular Edema (DME), is another retinal disease which is characterized by the accumulation of exudative fluid in the macula. DME is the most prevalent form of sight-impendent retinopathy in people with diabetes \cite{3}. The breakdown of the inner blood-retinal barrier is one of the reasons for this problem (see Fig. \ref{fig_1}). Early detection of retinal abnormalities is considerably important in preventing DME and subsequent loss of vision problem \cite{4}. 6\% of diabetic people are affected by DME which can resulted in more than 20 million cases worldwide \cite{3}. 

In the case of identification of AMD and DME, optical coherence tomography (OCT) could be the first contributory imaging modality. OCT as a noninvasive technology can demonstrate basic structural changes of the retina, such as retinal swelling, cystoid edema, serous retinal detachment, photoreceptor layer status and atrophy. Such a powerful technique can be used for diagnosing of macular edema even at early stages. Today, Spectral Domain (SD)-OCT which is the newer generation of OCT is developed in the field of rate of acquisition and resolution. These abilities of OCT follow with new options like eye tracking, 3D image segmentation and better imaging of the choroid. Providing more clear retina structural information and the ability to measure the thickness of its layers leads to a better understanding of pathology, the detection of AMD and DME, and the assessment of the effectiveness of treatment. \cite{5}.

Visualization of intraocular structures, such as macula and optic nerves is already possible by SD-OCT which measures the optical back scattering of the tissues. SD-OCT can visualize the internal structures of the retina and makes it possible to diagnose aforementioned diseases \cite{6}. 
Despite of the fact that OCT has gained significant importance in recent years for AMD and DME detection, the manual analysis remains time-consuming. Generally, 3D OCT volumes are comprised of 50-100 cross-sectional B-scans. Basically, investigation of each B-scan is necessary in the proper assessment of most retinal diseases. In addition to this time-consuming process, multiple cross-sections should be inspected simultaneously to identify scarce traces of early-stage ocular diseases. Considering these problems, automated algorithms for pathology identification in OCT volumes could be a great help for clinicians and ophthalmologists  \cite{7}. 

\begin{figure*}
	\centering{\includegraphics[scale=1]{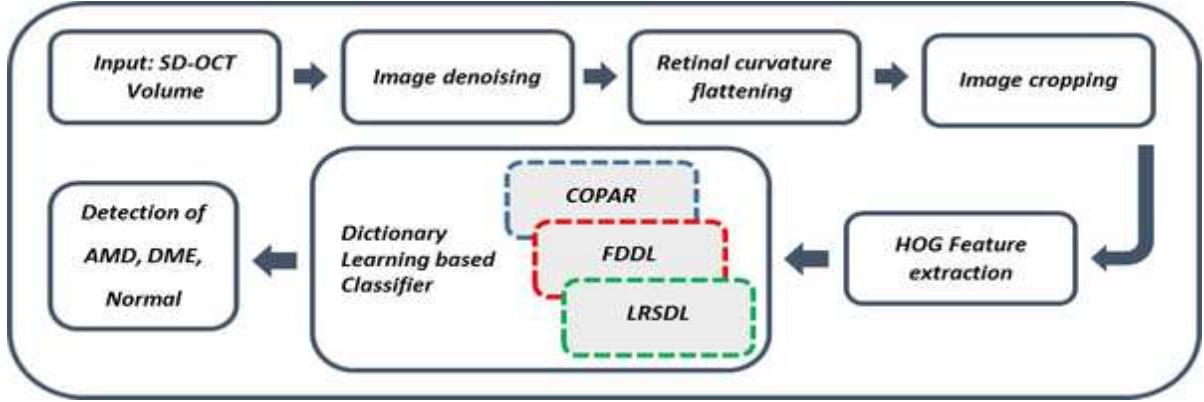}}
	\caption{Overview of the proposed approach\label{fig_2.eps}}
\end{figure*}

With more focus on specific to automatic pathology identification, various studies have explored detection of retinal pathologies using machine learning techniques, either focusing on segmentation of relevant pathological markers or retinal layers  \cite{8}. Despite of their considerable results, most of these works have leveraged B-scan level ground truth information. These more detailed labels are often not available or in some cases are not completely reliable, which can be a limitation for the usability of these solutions  \cite{7}.

There are some other studies for automatic identification of ocular diseases which are not focused on a direct comparison of retinal layer thicknesses. In \cite{8} kernel principal component analysis model (KPCA) ensembles were used for a binary classification of normal subjects and patients with advanced AMD. The author proposed to create ensembles by employing different features of each class and training one-class KPCA models. In order to combine KPCA models, a product combination rule was introduced. Finally, each new image could be classified based on its calculated confidence scores. The recognition rate of 92\% for each class of normal and AMD subjects was reported using this method \cite{9}. By the use of a similar data set, extraction of oriented gradient local binary pattern histogram was introduced in \cite{10}. A Bayesian network classifier was then performed on advanced AMD and normal subjects which resulted in an area under curve performance of 0.94. 

Beside many studies focused on the classification of AMD/normal or DME/normal cases \cite{8, 10, 11, 12}, there are three recent papers which are focused on the classification of AMD, DME and normal cases (3 class classification). For providing a complete comparison, in the following we introduce their methods and finally in section \ref{sec3} compare their results with ours. 

The first work is \cite{13} which uses Histogram of Oriented Gradients (HOG) descriptors and SVMs. The authors reported the accuracy of 100\% at detection of AMD and DME and 86.67\% for normal subjects’ identification. 

The second study \cite{14} has employed linear configuration pattern (LCP) features and used Correlation-based Feature Subset (CFS) selection algorithm as their representation tool. The accuracy of 100\% for both DME and normal samples and 93.33\% for AMD data was reported. 

In the latest similar study to our research \cite{15}, first a preprocessing step for aligning and cropping retinal regions is applied and then global representations of images based on sparse coding and a spatial pyramid are obtained. A multi-class linear support vector machine (SVM) classifier was used for classification. The success of this study was on completely correct identification of DME and AMD cases and achieving the accuracy of 93.33\% for normal subjects’ detection. 

OCT images of patients with DME and AMD have common parts with images of normal people and some parts which are specific to the diseases. Dictionary learning methods separate image to sub-particles which some of them are common and some others are discriminative. It seems that dictionary learning is an approach that mimics our understanding by learning common and discriminative patterns in OCT datasets. So we propose to use dictionary learning methods to detect and classify retina diseases from OCT images. In this paper,  utilization of HOG features of pyramid images in conjunction with 3 different dictionary learning methods including separating the Particularity and the Commonality dictionary learning (COPAR), Fisher Discrimination Dictionary Learning (FDDL) and Low-Rank Shared Dictionary Learning (LRSDL), were investigated to provide the highest classification accuracy of OCT images. 

The remainder of the paper is organized as follows.  Section \ref{sec2} introduces our proposed method which consists of 6 main parts: subsections \ref{subsec2.1}-\ref{subsec2.4} represent the preprocessing steps and feature extraction procedure and subsections \ref{subsec2.5}-\ref{subsec2.6}  provide an explanation of different classifier training based on dictionary learning methods. In Section \ref{sec3} we demonstrate our experimental results through the leave-three-out cross-validation, and Section \ref{sec4} outlines conclusion and summary.

\section{Method}\label{sec2}

In this section, we introduce our proposed method for identification of retinal abnormalities based on classifying SD-OCT B-Scans. The main pipeline of this method includes image denoising, flattening the retinal curvature, cropping, HOG feature extraction and classification using dictionary learning method. This procedure is summarized in Fig.~\ref{fig_2}. Details of each step will be clarified in the following subsections.

\subsection{Image denoising}\label{subsec2.1}

OCT imaging technique like many other imaging modalities due to instrument limitations and unwanted artifacts provides noisy and low visual contrast images.  So at the first step of working with OCT images, a contrast enhancement step is recommended. To achieve this goal, we used the proposed method in \cite{16,16_1} which is specifically focused on denoising OCT images. The authors consider the layered structure of retina in OCT images with a specific probability distribution function (pdf) diminished by speckle noise. They proposed to employ statistical modeling of OCT images by the means of a mixture model. By considering the Normal-Laplace distribution, each component of this mixture model would be created. The proposed distribution is based on the convolution of Laplace pdf and Gaussian noise. The observed decaying behavior of OCT intensities in each layer of normal OCTs is the main reason for selecting Laplace pdf. After fitting a Normal-Laplace mixture model to the data, each component is Gaussianized and all of the components are combined with Averaged Maximum A Posterior (AMAP) method \cite{16_2}. After all, they proposed a new contrast enhancement method based on their proposed statistical model. Considering the fact that OCT images suffer from speckle noise as a multiplicative noise, author suggested to use a logarithm operator to convert this noise to additive Gaussian noise. Therefore, first OCT images are transformed to the logarithmic domain and then all the processes are implemented in this domain. The second step is fitting a Normal-Laplace mixture model to the OCT data by using Expectation Maximization (EM) algorithm for estimation of its parameter. Each component of the estimated mixture model has transformed to a Gaussian distribution with specific mean and variance so that each enhanced component could be obtained. In order to construct the complete enhanced image, the gaussianized components are combined with each other. For this combination, a weighted summation is used which is similar to the AMAP method. Finally, by applying the exponential operator to the combined component, the enhanced data would be obtained. This enhancement method in addition to normal images for non-healthy cases also has good results.

\subsection{Retinal curvature flattening}\label{subsec2.2}

SD-OCT images of the retina have a curvature which varies both between patients and within each volume. One part of this curvature is due to natural anatomy of eyes and the other part is related to OCT image acquisition. In order to provide a similar platform for all images and to eliminate the effect of retina curvature during the classification, flattening the curvature of retina images is recommended. As described in \cite{13}, a primitive estimate of the RPE layer is calculated in the first step. Using the hyper-reflective characteristic of RPE layer, two highest maxima in each column are selected as RPE location candidates and then the outer most of them is chosen as the final estimation. In order to eliminate outlier points using median filter is suggested. To achieve lower boundary of the retina, we performed convex hull of the primitive estimation of RPE points, and then consider its lower border as the retina boundary. To achieve a smoother curve, a further step of fitting a polynomial to the RPE estimated boundary also could be done. According to the visually estimated curvature of each image, degrees of polynomial differs from 1 to 4.  Finally, the flattened retinal boundary is obtained by shifting each column up or down so that the points of the estimated curve are placed on a flat line. 

In some cases that the results of this method is not good enough, we suggested another approach. Instead of finding the convex hull, we focused on the primitive estimated points of the RPE layer. Because some of these points were far from the lower bound of the RPE, we calculated the vertical distance between the lowest and highest points and only considered a fraction of this distance and its inclusion points. 
This caused to points which were more representative of the lower boundary of the RPE. Fitting a polynomial to these points could result in a better-flattened image. 

\begin{figure}[!t]
	\centering
	\begin{subfigure}[b]{0.45\linewidth}
		\includegraphics[width=\linewidth]{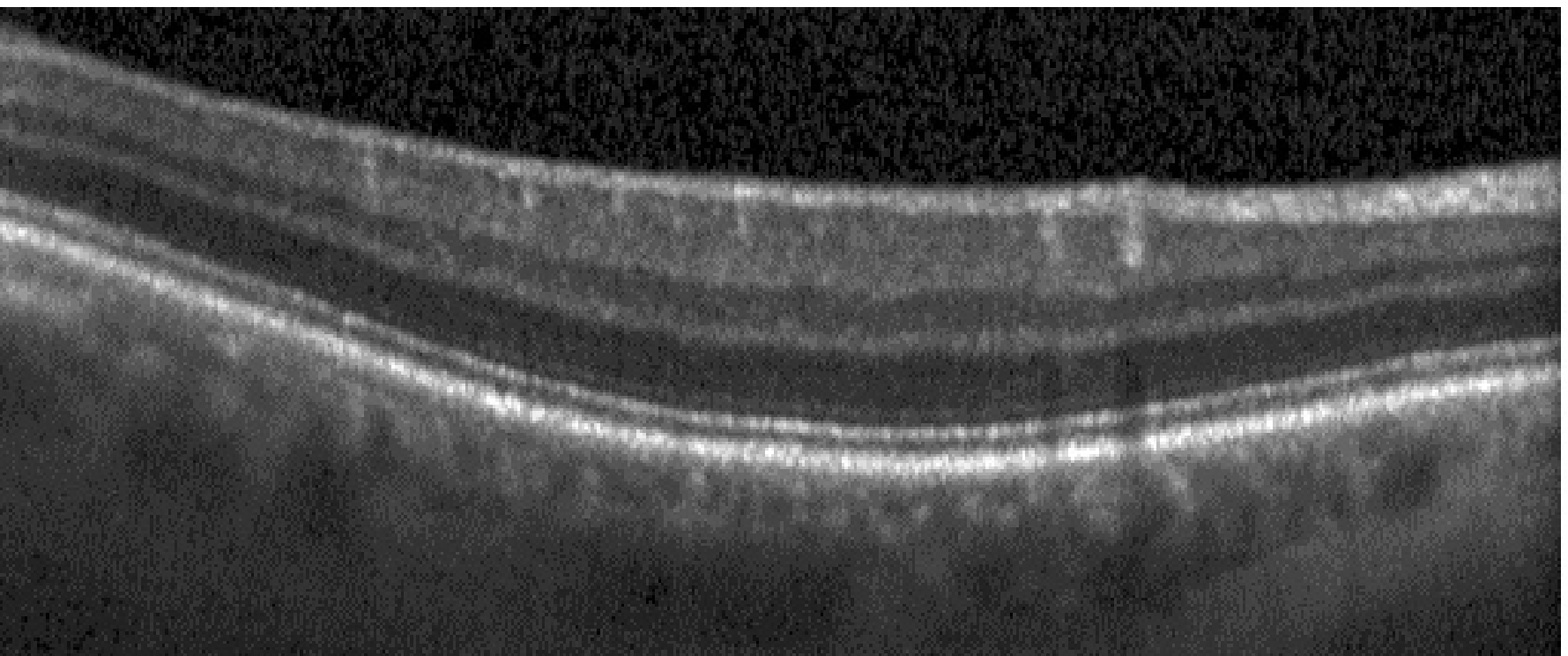}
		\caption{}
		\label{fig:Norm1}
	\end{subfigure}
	~ 
	\begin{subfigure}[b]{0.45\linewidth}
		\includegraphics[width=\linewidth]{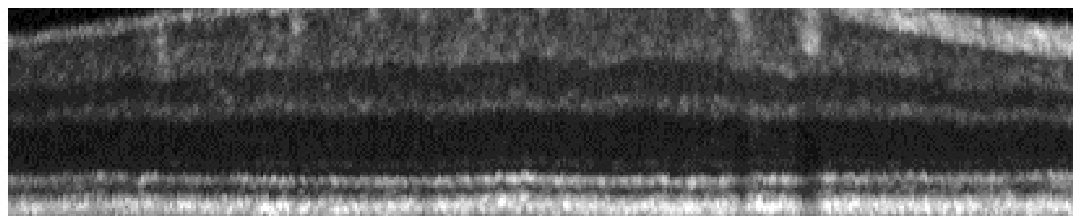}
		\caption{}
		\label{fig:Norm2}
	\end{subfigure}
	~ 
	\begin{subfigure}[b]{0.45\linewidth}
		\includegraphics[width=\linewidth]{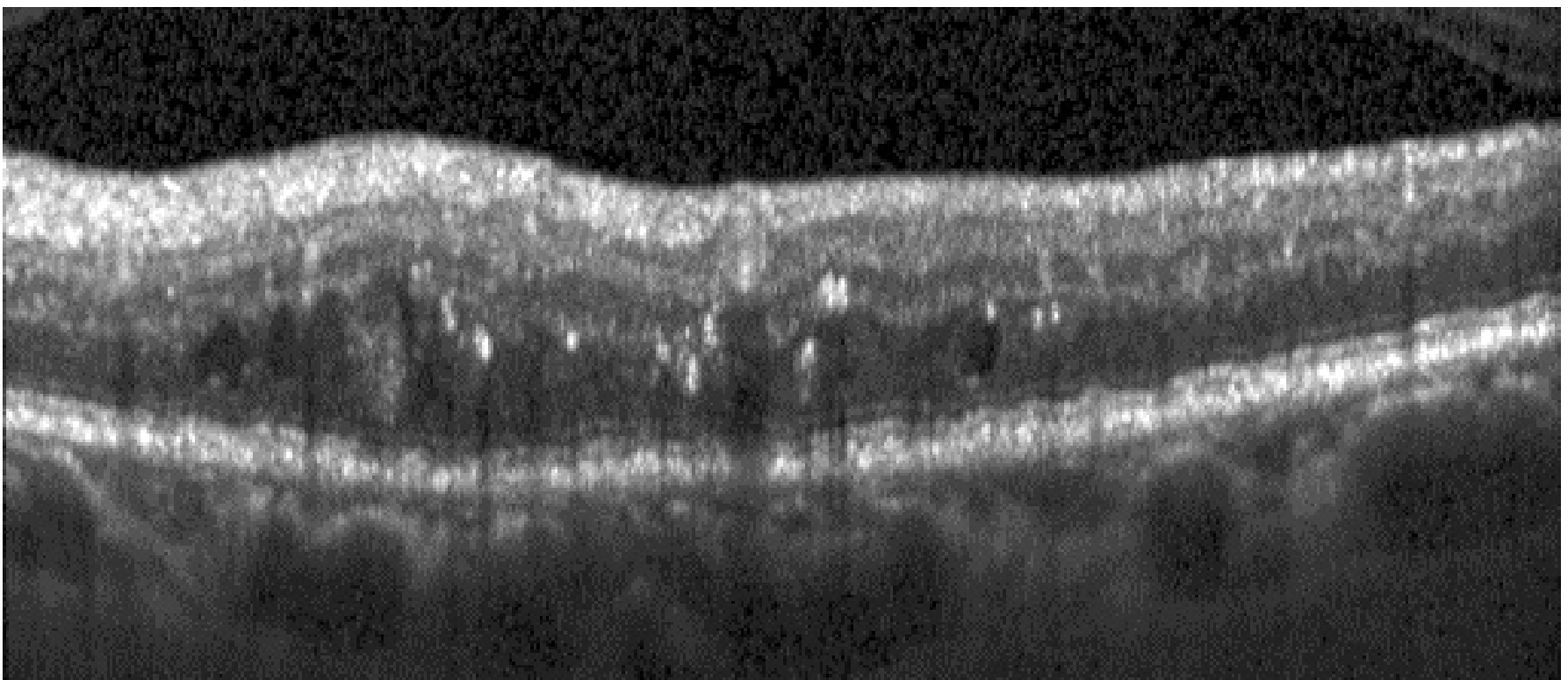}
		\caption{}
		\label{fig:DME1}
	\end{subfigure}
	\begin{subfigure}[b]{0.45\linewidth}
		\includegraphics[width=\linewidth]{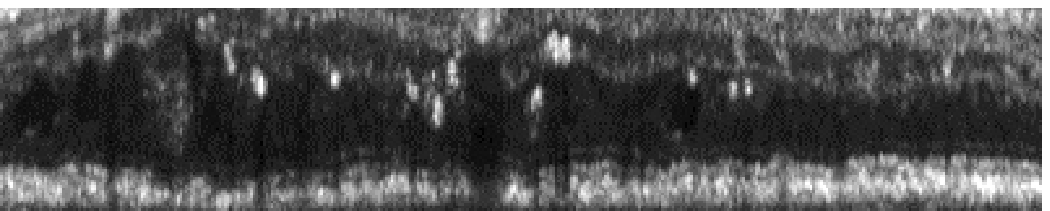}
		\caption{}
		\label{fig:DME2}
	\end{subfigure}
	\begin{subfigure}[b]{0.45\linewidth}
		\includegraphics[width=\linewidth]{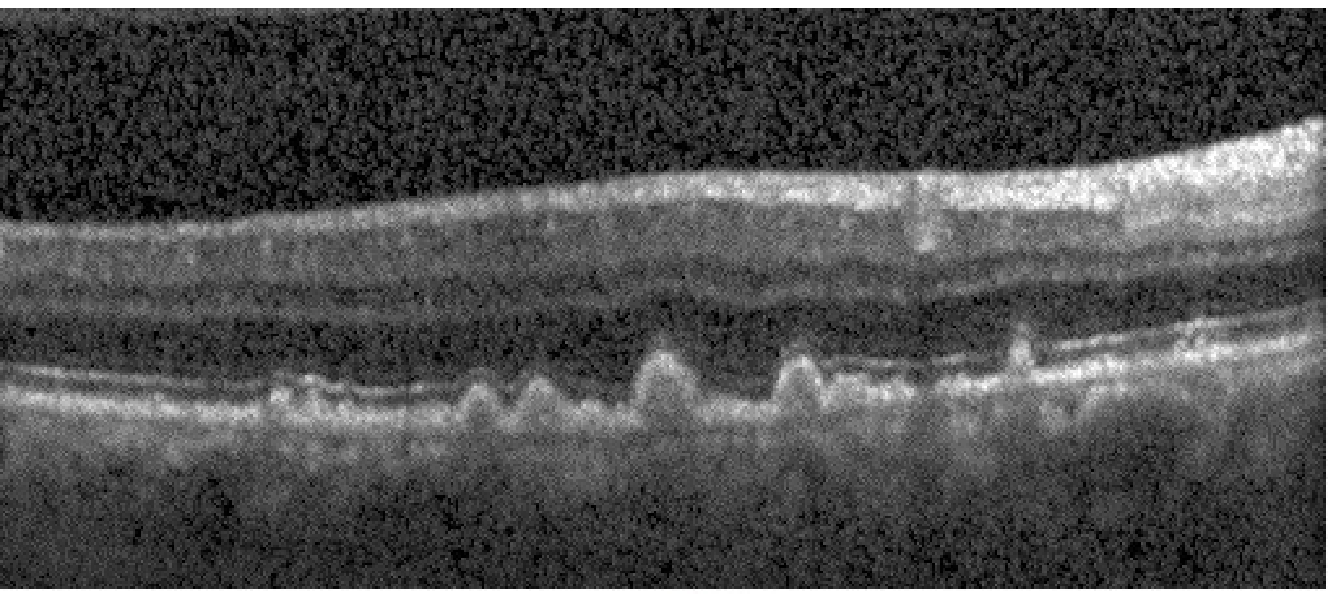}
		\caption{}
		\label{fig:AMD1}
	\end{subfigure}
	\begin{subfigure}[b]{0.45\linewidth}
		\includegraphics[width=\linewidth]{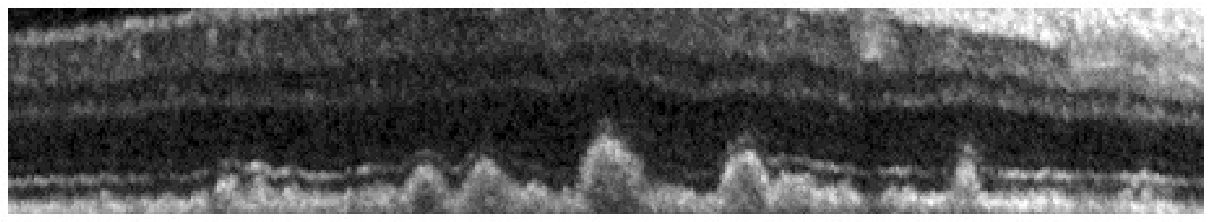}
		\caption{}
		\label{fig:AMD2}
	\end{subfigure}
	\caption{Samples of SD-OCT images, left column includes a) Normal, c) DME and e) AMD B-scans, respectively, and right column (b,d,f) includes their corresponding denoised, flattened and cropped images. }\label{fig_3}
\end{figure}

\subsection{Image cropping}\label{subsec2.3}

In our study, the area under investigation is constrained to a window in the vicinity of the fovea. Its lower bound is limited to RPE layer while neglecting choroid. In order to take into account the changes of various diseases and providing the same dimension for feature extraction step, we crop all SD-OCT images in the same size. While the width of each SD-OCT cropped image is considered to be 380 pixels, each image consists of 65 pixels in the axial direction, nearly 60 above and 5 below the estimated RPE. Fig \ref{fig_3} demonstrates three B-scans and their denoised flattened and cropped versions. 

\subsection{Feature extraction }\label{subsec2.4}

One of the main steps of a good classification is choosing the best feature with regard to the application. As suggested in \cite{13}, HOG descriptors are powerful tools for discriminating normal and abnormal OCT images. HOG descriptor \cite{16_3}, counts the number of gradient orientation incidences in the partitions of an image . Although this method is similar to edge orientation histograms, feature transform descriptors and shape contexts, the main difference refers to the step of computation which use a dense grid of uniformly spaced partitions. The basic step of HOG descriptor calculation is partitioning the image into small sub images called cells, and larger regions called block containing some cells while blocks can overlap with each other. By calculating the histogram of gradient of each cell, normalizing them on each block and concatenating them to each other, computation of HOG descriptors could be completed. The main reason for normalizing is improving accuracy and illumination invariance. In this paper we used the same procedure as \cite{13} for the feature extraction step and focused on dictionary learning methods for classification step. We calculated the HOG descriptors of Gaussian image pyramid. Creating an image pyramid consists of two main step: smoothing and subsampling of the smoothed image. In this paper, smoothing has done based on a separable kernel $\mathbf{W} (m, n) = \mathbf{w} (m) \mathbf{w} (n)$ with $\mathbf{w} = [(1/4) - (a/2), 1/4, a, 1/4, (1/4) - (a/2)]$. In order to form the weighting function close to a Gaussian, we set the value of the parameter $a$ equal to 0.375. Based on different experiments the best of our results achieved with the cell size $4\times4$, constructed block using $2\times2$ cells and block overlap of $1\times1$. Finally, we use the normalized histograms of all the blocks and construct a feature vector for each image. 

\subsection{Sparse representation and dictionary learning framework}\label{subsec2.5}

It is shown that most of signals and images can be expressed by a linear combination of a few bases taken from a  “dictionary”. This concept refers us to the sparse representation-based classifier (SRC) which is already developed and adapted to numerous signal/image classification task. So, based on this theory, a dictionary is learned from training samples instead of using all of them as prior information and try to represent new images using the learned dictionary. 

The main aim of sparse representation is to represent a given signal $\mathbf{y}$  of dimension $n$ as a linear combination of a small number of bases derived from a collection of extracted features which is known as a dictionary. Each column of a dictionary is called an atom. 
We show the dictionary by $\mathbf{D}$ and its atoms of size $n$  by $\mathbf{d}_k$ where $k=1, \ldots, N$ and $N$ is the number of columns of the dictionary. So each matrix $\mathbf{D} \in \mathbb{R}^{n \times N}$ is constructed by column concatenating of $N$ vectors $\mathbf{d}_k$. If $N\geq n$, the dictionary is over complete and there is a dependency among its atoms. In that case, every signal $\mathbf{y}$ can be represented as a linear combination of atoms in the dictionary:
\begin{equation}\label{e1} 
\mathbf{y}= \mathbf{Dx} = \sum\limits_{k = 1}^N {{x}^k \mathbf{d}_k}  
\end{equation} 
where $\mathbf{x}$ is the coefficient vector and ${x}^k$ shows its elements. 

Considering the fact that the dictionary is over complete, $\mathbf{x}$ is not unique and so sparsity constraint can be mentioned. Obtaining a sparse representation of $\mathbf{y}$ could be done based on achieving a sparse vector $\mathbf{x}$ that most of its elements are close or equal to zero except a small number of significant coefficients. Suppose that $\mathbf{D}$ is constructed directly by column concatenation of training samples of different classes. In that case, the task of sparse representation of a signal can be  shown as the following optimization problem: 
\begin{equation}\label{e2} 
\mathop {\arg\min \limits_{\mathbf{x}} (||\mathbf{y} - \mathbf{Dx}|| _2^2 + \lambda ||\mathbf{x}||_1)} 
\end{equation} 
where $\lambda$ is a constant for weighting the sparsity constraint. 

At the first step of dictionary learning formulation, it could be an extension of sparse representation of training samples, so that dictionary should be learned during the optimization of (\ref{e2}). At the second step, this formulation could be expanded so that the dictionary becomes simultaneously reconstructive and discriminative. Utilization of an additive discriminative term based on labels of training data in the dictionary learning methods could guarantee that the data representation of each class is sufficiently different. So, learning a dictionary as a supervised problem minimizes the sparse approximation errors over different classes while imposes discrimination between classes \cite{2}. In order to achieve these goals, the general dictionary learning problem can be formulated as below:

\begin{equation}\label{e3}
\mathop {\arg\min\limits_\mathbf{X,D} (||\mathbf{Y} - \mathbf{DX}||_2^2}  + {\lambda _1}||\mathbf{X}||_1 + {\lambda _2}h(\mathbf{D,X,\theta} ))
\end{equation}
where $\mathbf{Y}$ and $\mathbf{X}$ show the set of training samples and their corresponding sparse coefficient matrix respectively and $h(\mathbf{D,X,\theta} )$ is the discriminative term which could be developed based on different methods. 

It has been proved that, when different algorithms based on different cost functions and different discriminative terms are used on a specific class of signals, some dictionaries provide a better approximation performance. In other words, there exist dictionaries that provide a more acceptable sparse solutions \cite{17}. 
In order to classify OCT images of three aforementioned classes using the HOG features, we investigated three different dictionary learning algorithms and based on a comparison among their results, we introduce one of them as the best classifier for the proposed pipeline. 

\begin{figure*}
	\centering{\includegraphics[scale=1]{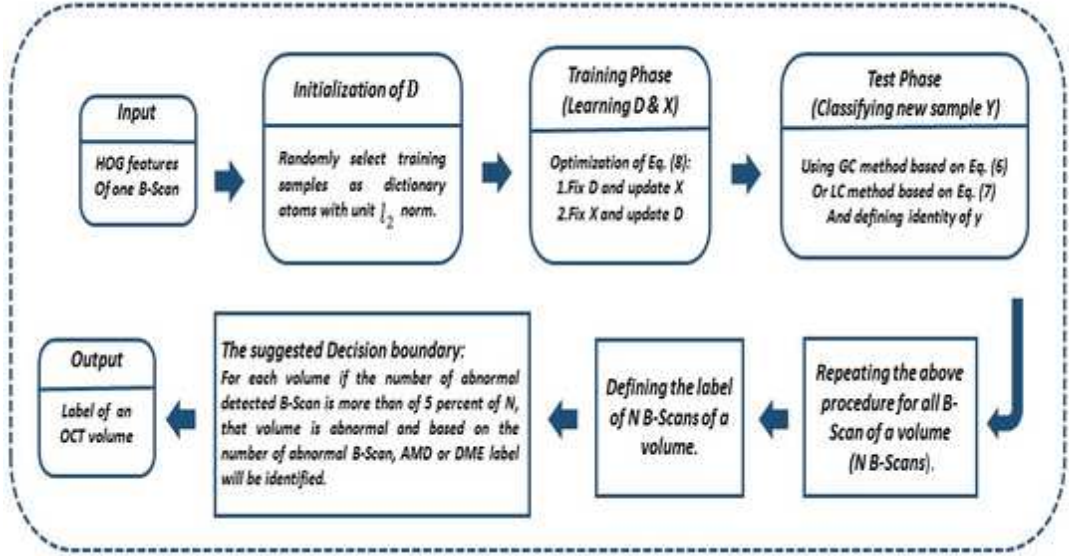}}
	\caption{Classification of one SD-OCT volume using FDDL algorithm\label{fig_4}}
\end{figure*}

\subsection{Image classification}\label{subsec2.6}

\subsubsection{COPAR}\label{subsubsec2.6.1}

The first used algorithm is particularity and the commonality dictionary learning (COPAR) \cite{18}. This algorithm tries to explicitly learn the most discriminative features of each category (the particularity) and some common feature observed in all the training set (the commonality). The particularity part of the dictionary is related to specific atoms of each class and the commonality corresponds to shared atoms of all classes. Shared atoms are the representative features of all categories and essential for reconstruction of all the data. The contribution of adding commonality dictionary provides more comprehensive information for the task of classification. In order to investigate deeper, we review the basic of this dictionary learning method based on its cost functions.

In general, to learn a dictionary with purpose of classification, we can learn $C$ class-specific dictionaries  ${\mathbf{D}_c}$’s, for all $C$ classes $(c = 1,\ldots, C)$. In the ideal case, it is expected that each object would be represented by its own class subspace and there is no overlap among all class subspaces. In another word, given $C$ classes and a dictionary $\mathbf{D} = [{\mathbf{D}_1},\ldots,{\mathbf{D}_C}]$ with ${\mathbf{D}_c}$ representing training samples from class $c$, a new sample $\mathbf{y}$ from class $c$ can be represented as $\mathbf{y} \approx \mathbf{D}_c \mathbf{X}_c$. Therefore, if we express $\mathbf{y}$ using the dictionary $\mathbf{D}:\mathbf{y} \approx \mathbf{DX} = {\mathbf{D}_1}{\mathbf{X}_1} + \ldots + {\mathbf{D}_c}{\mathbf{X}_c} + ...{\mathbf{D}_C}{\mathbf{X}_C}$, then the more effective elements of $\mathbf{X}$ should be located in ${\mathbf{X}_c}$ and hence, the coefficient $\mathbf{X}$ should be sparse. In matrix form, if we show samples of class $c$ by ${\mathbf{Y}_c}$, then $\mathbf{Y} = [{\mathbf{Y}_1},...,{\mathbf{Y}_c},...,{\mathbf{Y}_C}]$ is the set of all samples and the coefficient matrix $\mathbf{X}$ would be sparse. The class-specific dictionaries  ${\mathbf{D}_c}$’s from different classes usually describe some common atoms, which are not useful for classification but are essential for the reconstruction step of dictionary learning. Although existence of these common atoms reduce the performance of classification, but they are essential for representation of a query datum. Regarding this problem, to improve the classification performance, COPAR algorithm explicitly separate the coherent atoms by learning the commonality ${\mathbf{D}_0}$, which provides the common bases for all classes. So, the overall dictionary form as  $\bar{\mathbf{D}}=[\mathbf{D},\mathbf{D}_0] \in \mathbb{R}^{d\times K}$ in which $K = \sum\nolimits_{c = 0}^C {{K_c}} $ and ${K_c}$ is the number of atoms of each class, $\mathbf{D}_c \in \mathbb{R}^{d\times K_c}$ stands for the particularity of ${c^{th}}$ class, and $\mathbf{D}_0 \in \mathbb{R}^{d\times K_{0} }$  is the commonality. 
In order to provide the best representation of samples of each class, we use the joint particular dictionary of that class and the commonality dictionary. Denoting by $\mathbf{X}^i$ the sparse coefficient of $\mathbf{Y}$ on ${\mathbf{D}_i}$, by $\mathbf{X}_c \in \mathbb{R}^{K\times N_c}$ the sparse coefficient of $\mathbf{Y}_c$ on $\mathbf{D}$, by $\mathbf{X}_c^i$ the sparse coefficient of $\mathbf{Y}_c$ on ${\mathbf{D}_i}$, the total sparse coefficient could be represented by $\overline{\mathbf{X}} = [\mathbf{X}^T , (\mathbf{X}^0)^T]^T$. Considering the above mentioned notation, a learned dictionary $\mathbf{D}$ should be able to well represent every sample ${\mathbf{Y}_c}$, i.e. ${\mathbf{Y}_c} \approx \mathbf{D}{\mathbf{X}_c}$. In addition to the overall dictionary, it is also expected that the sample from the ${c^{th}}$ class could be well represented by the cooperative efforts of the ${c^{th}}$ particular dictionary ${\mathbf{D}_c}$ and the shared dictionary ${\mathbf{D}_0}$. Therefore COPAR objective function $f$ has been introduced as: 

\begin{equation}\label{e4}
f \equiv \frac{1}{2}g(\mathbf{Y},\overline{\mathbf{D}} ,\overline{\mathbf{X}} ) + \lambda_{1}||\overline{\mathbf{X}} ||_{1} + \frac{{\lambda_{2}}}{2}h(\mathbf{D})
\end{equation}
where $g(\mathbf{Y},\overline{\mathbf{D}} ,\overline{\mathbf{X}} )$ is defined as: 
\begin{equation}\label{e44}
\begin{array}{l} 
\sum\limits_{c = 1}^C \{ { ||{\mathbf{Y}_c} - \overline{\mathbf{D}}\overline{\mathbf{X}}_c||_F^2}  + ||{\mathbf{Y}_c} - {\mathbf{D}_0}\overline{\mathbf{X}}_c^0 - {\mathbf{D}_C}\overline{\mathbf{X}}_c^c||\} \\ 
+ \sum\limits_{j = 1,j \ne c}^C {||\overline{\mathbf{X}}_c^j||_F^2} 
\end{array}
\end{equation}

Simultaneously usage of particularity and commonality dictionaries provide the comprehensive representation of samples of each category which is placed in $g(\mathbf{Y}, \overline{\mathbf{D}}, \overline{\mathbf{X}})$ by its two first terms. It is worth mentioning that the estimated sub-dictionaries based on only the two first terms may contain atoms which use interchangeably. So, in order to provide incoherency between sub-dictionaries, the third term is imposed in $ g(\mathbf{Y}, \overline{\mathbf{D}}, \overline{\mathbf{X}})$. 

In addition to incoherency of class-specific sub-dictionaries, $h(\mathbf{D}) = \sum\limits_{c = 0}^C {\sum\limits_{i = 0,i \ne 0}^C {||\mathbf{D}_i^T{\mathbf{D}_c}||_F^2}}$, the last penalty term of  (\ref{e4}) has been used to enforce the incoherency among the commonality with the particularities. Finally by minimizing  (\ref{e1}), $\mathbf{D}, {\mathbf{D}_0}, \mathbf{X}$ and ${\mathbf{X}^0}$ would be calculated. 

In classification step, authors propose to adopt two reconstruction errors based on selected classification schemes: Global classifier (GC) and Local classifier (LC): 

\textbf{GC}: When the number of training samples of each class is relatively small, the learned sub-dictionary ${\mathbf{D}_i}$ may not be able to provide representative samples of ${i^{th}}$ class. So, we refer to the whole dictionary $\overline{\mathbf{D}}$ for representing the query sample $\mathbf{y}$.
In order to increase the speed of computation in the test stage, the $l1$-norm regularization on the representation coefficient could be relaxed to $l2$-norm regularization. With these considerations, the following global representation mode introduced which code a testing sample 
$\mathbf{y}$ over the learned dictionary $\overline{\mathbf{D}}$:

\begin{equation}\label{e5}
e = \mathop {\arg \min }\limits_\mathbf{X} ||\mathbf{y} - \overline{\mathbf{D}} \overline{\mathbf{X}} ||_2^2 + \gamma ||\overline{\mathbf{X}}||_{1}
\end{equation}
where $\gamma$  is a constant.

\textbf{LC}:
By increasing the number of training sample of each class, the representation power of each sub space by its sub-dictionary would be higher. In this case, to overcome the problem of interference among sub-dictionaries it is suggested to represent a testing sample $\mathbf{y}$ locally over each sub-dictionary $\mathbf{D}_c$  instead of the whole dictionary $\mathbf{D}$. So we calculate $C$ reconstruction error for the $C$ classes as below:

\begin{equation}\label{e6}
{e_c} = \mathop {\arg\min }\limits_{{{\overline{\mathbf{X}} }_c}} ||y - {\overline{\mathbf{D}} _c}{\overline{\mathbf{X}} _c}||_2^2 + \gamma ||\overline{\mathbf{X}} _c||_{1}
\end{equation}
The final diagnosed label of new sample $\mathbf{y}$ is $\overline{c}  = \mathop{\arg \min }\limits_c {e_c}$.

\subsubsection{FDDL}\label{subsubsec2.6.2} 

The next used algorithm is Fisher Discrimination Dictionary Learning (FDDL). In this algorithm a dictionary could be learned based on the Fisher discrimination criterion. The main effort in FDDL is toward learning a structured dictionary, whose atoms have correspondences to the subject class labels. In order to achieve this aim, dictionary is not only based on the representation residual to recognize different classes, but also consider the representation coefficients to be discriminative. Based on the Fisher discrimination criterion, $\mathbf{X}$ could be achieved by minimizing within-class scatter and maximizing between class scatter \cite{19}. Using above-mentioned notation and concepts, FDDL cost function can be formulated as below: 

\begin{equation} \label{e7}
f \equiv \frac{1}{2}{g}(\mathbf{Y,D,X}) + {\lambda _1}||\mathbf{X}||_{1} + \frac{\lambda _2}{2}h(\mathbf{X})
\end{equation}

The discriminative fidelity term of FDD, $g(\mathbf{Y,D,X})$, focused on the representation residual, is defined as follow: 

\begin{equation}\label{e8}
\begin{array}{l} 
g(\mathbf{Y,D,X}) = \frac{1}{2}\sum\limits_{c = 1}^C \{ ||\mathbf{Y}_c  - \mathbf{D}\mathbf{X}_c||_F^2  + \\ 
||\mathbf{Y}_c - \mathbf{D}_c \mathbf{X}_c^c||_F^2 \}  + \sum\limits_{j = 1,j \ne c}^C ||\mathbf{D}_j \mathbf{X}_c^j||_F^2 
\end{array}
\end{equation}

The first penalty term only relies on $\mathbf{D}$ to represent $\mathbf{Y}_c$ while its correspondence residual may digress much from $\mathbf{Y}_c$. So, $\mathbf{D}_i$ would not be able to well represent the $\mathbf{Y}_c$. Adding the second penalty term is the suggested solution to avoid this problem. In spite of that, it is possible that other sub-dictionaries could cooperate in representing $\mathbf{Y}_c$ which reduce the discrimination power of $\mathbf{D}_c$. The responsibility of the third term is reducing the representation power of $\mathbf{D}_j$ to $\mathbf{Y}_c$, $c \neq j$.  

In order to increase the discriminative power of dictionary, another constraint is imposed toward discrimination of sparse coefficient. 
The Fisher-based discriminative coefficient term,  which its main idea is minimizing the within class scatter and maximizing the between class scatter is defined as follow:

\begin{equation}\label{e9}
h(\mathbf{X}) = \sum\limits_{c = 1}^C {\{ ||{\mathbf{X}_c} - {\mathbf{M}_c}} ||_F^2 - ||{\mathbf{M}_c} - \mathbf{M}||_F^2\} + ||\mathbf{X}||_F^2
\end{equation}

If we define $\mathbf{m}$ and ${\mathbf{m}_c}$ as the mean value of $\mathbf{X}$ and $\mathbf{X}_c$ columns, respectively, $\mathbf{M}_c =[\mathbf{m}_c, ... , \mathbf{m}_c]\in \mathbb{R}^{K\times N_c}$, $\mathbf{M} = [\mathbf{m},...,\mathbf{m}]$ are the mean matrices. The number of columns of $\mathbf{M}$ depends on context, e.g. by writing $\mathbf{M} -\mathbf{M}_{c}$, we mean that $\mathbf{M}$ and $\mathbf{M}_{c}$ have same number of columns ${N_c}$.

In the classification step of FDDL algorithm for new test samples, we use GC and LC methods as described for COPAR method. A summary of using FFDL consist of learning phase and classification step of extracted features is shown in Fig. \ref{fig_4}. 

\subsubsection{LRSDL}\label{subsubsec2.6.3}

The last used dictionary learning method with the aim of SD-OCT data classification is Low Rank Shared Dictionary Learning (LRSDL) which is inspired by COPAR algorithm \cite{20}. The authors propose a method to simultaneously and explicitly learn a set of common structures as well as class-specific features for classification.  By considering the importance of the shared bases, authors suggested enforcing two constraints on the shared dictionary. As the shared subspace gets wider, the shared atoms may also get interference with discriminative atoms. So to circumvent this problem, the low-rank constraint should be imposed on the shard dictionary. A low-rank structure results in a low dimensional spanning subspace which only includes the most common features. Authors point out that contribution of the shared dictionary in every signal reconstruction should be close together. Based on this idea the second constraint focused on sparse coefficients so that sparse coefficients corresponding to the shared dictionary should be almost similar. Regarding all the above condition for learning the shared dictionary, LRSDL cost function is summarized as:

\begin{equation}\label{e10}
f \equiv \frac{1}{2}g(\mathbf{Y}, \overline{\mathbf{D}} ,\overline{\mathbf{X}} ) + \lambda_{1}||\overline{\mathbf{X}} ||_{1} + \frac{\lambda _{2}}{2}h(\overline{\mathbf{X}} ) + \eta ||{\mathbf{D}^0}|{|_*}
\end{equation} 
where $g(\mathbf{Y}, \overline{\mathbf{D}} ,\overline{\mathbf{X}}) $ tries to well represent $\mathbf{Y}_c$ by collaboration of the particular dictionary ${\mathbf{D}_c}$ and the shared dictionary ${\mathbf{D}_0}$. By employing the discriminative fidelity term of (\ref{e8}) and considering the cooperation of ${\mathbf{D}_0}$ and ${\mathbf{D}_c}$, $g(\mathbf{Y}, \overline{\mathbf{D}} ,\overline{\mathbf{X}} )$ could be rewritten as:

\begin{equation}\label{e11}
\begin{array}{l}
g(\mathbf{Y}, \overline{\mathbf{D}} ,\overline{\mathbf{X}})  = \sum\limits_{c = 1}^C {\{ ||{\mathbf{Y}_c}- \overline{\mathbf{D}} \overline{\mathbf{X}}_{c}||_F^2} \\ + ||{\mathbf{Y}_c} - {\mathbf{D}_0}\mathbf{X}_c^0 - {\mathbf{D}_c}\mathbf{X}_c^c||_F^2\} + \sum\limits_{j = 1,j \ne c}^C {||\mathbf{D}_j^T\mathbf{X}_c^j||_F^2}  
\end{array}
\end{equation}

Considering the above-mentioned point, Fisher-based discriminative coefficient term of (\ref{e10}), appears as: 

\begin{equation}\label{e12}
\begin{array}{l}
h(\overline{\mathbf{X}}) = \sum\limits_{c = 1}^C {\{ ||{\mathbf{X}_c} - {\mathbf{M}_c}||_F^2 - ||{\mathbf{M}_c} - \mathbf{M}||_F^2\} }
+ ||\mathbf{X}||_F^2\\+ ||{\mathbf{X}^0} - {\mathbf{M}^0}||_F^2 
\end{array}
\end{equation}

The only difference between (\ref{e9}) and (\ref{e12}) is the existence of term $||\mathbf{X}^0 -\mathbf{M}^0||^2_F$ which is due to the shared dictionary and its corresponding sparse coefficients. This term forces the coefficients of all training samples represented via the shared dictionary to be similar. 

Furthermore, for the shared dictionary, the author proposes minimizing $\mathbf{rank}\left( {\mathbf{D}_0} \right)$ which its convex relaxation is equivalent to the nuclear norm $||\mathbf{D}_{0}||_{*}$. After the learning process, which includes calculation of $\overline{\mathbf{D}}$, mean vectors $\mathbf{m}_{c}$ and $\mathbf{m}^{0}$, we can make a classification scheme for test samples. For a new test sample $\mathbf{y}$, authors suggested to find its coefficient vector $\mathbf{X} = [ \mathbf{X}^T, ( \mathbf{X}^0  )^{T}]^T$ with the sparsity constraint on $\overline{\mathbf{X}}$ and also encourage $\mathbf{X}^{0}$ to be close to $\mathbf{m}^{0}$:

\begin{equation}\label{e13}
\overline{\mathbf{X}}= \mathop{\mathop{\arg \min }\limits_{\overline{\mathbf{X}}}} ||\mathbf{y} - \overline{\mathbf{D}}\overline{\mathbf{X}} ||_2^2 + \lambda _{1}|| \overline{\mathbf{X}}||_{1} + \frac{\lambda _{2}}{2}||\mathbf{X}^{0} - \mathbf{m}^0||_2^2
\end{equation}

Using calculated $\overline{\mathbf{X}}$ and eliminating the contribution of the shared dictionary by $\overline{\mathbf{y}}  = \mathbf{y} - \mathbf{D}_{0}\mathbf{X}^{0}$, the identity of $\mathbf{y}$ is determined by: 
\begin{equation}\label{e14}
\mathop{\arg \min }\limits_{1 \le c \le C} (w||\mathbf{y} - \mathbf{D}_{c}\mathbf{X}^{c}||_2^2 + (1 - w)||\mathbf{X} - \mathbf{m}_{c}||_2^2)
\end{equation}
where $w$ is responsible for weighting the terms according to the problem.

\subsubsection{Comparison of COPAR, FDDL and LRSDL}\label{2.6.4}

The 3 discussed dictionary learning methods are among the strongest methods presented in recent years in the field of dictionary learning based classification. As it was mentioned earlier in spite of the existence of commonality among these 3 methods, there are also some subtle differences that have been applied to improve the calculation of the dictionary. COPAR algorithm focuses on learning a shared dictionary and a set of class-specific sub-dictionaries, while FDDL uses fisher discriminant criteria and only learns a structured dictionary which consists of a set of class-specific sub-dictionaries. LRSDL is a composition of COPAR and FDDL algorithm. It uses the idea of the shared dictionary of COPAR and also fisher discriminant criteria of FDDL while also imposes two important constraints on the shared dictionary. 
Although three above mentioned cost functions are not jointly convex to $(\mathbf{D, X})$, they are convex with respect to each of $\mathbf{D}$ and $\mathbf{X}$, when the other is fixed. In order to provide a strong platform for classification of OCT images, we compare the result of our algorithm based on these 3 methods in Section \ref{sec3}. 


\subsection{SD-OCT data sets }\label{subsec2.77}

The investigated SD-OCT data sets of this study comprise volumetric scans acquired from 45 patients: 15 normal subject, 15 patients with dry AMD, and 15 patients with DME \cite{13}. These data set were acquired under supervision of Spectralis SD-OCT (Heidelberg Engineering Inc., Heidelberg, Germany) imaging at Duke University, Harvard University, and the University of Michigan. Scanning protocol of the data set differs in lateral and azimuthal resolution but is the same in axial direction. Number of A-scans and B-scans also varies in the data set. In order to attain more information, we refer reader to \cite{13}. 

\section{Experimental results}\label{sec3}

The main goal of this study is finding an approach for automatic classifying of three different classes of OCT images, including normal subject, DME and AMD patients. We proposed the above-mentioned method which was composed of investigation of different kinds of dictionary learning-based classifier: COPAR, FDDL and LRSDL.

To affirm the results of each one of these classifiers, we performed a series of experiments. Evaluating the performance of different classifiers with regard to different number of training samples was the main core of the experiments. In 3 stages, the accuracy of each classifier was investigated which are as follow: 
\begin{itemize}
	\item  Training the classifiers based on 5 OCT volumes of each class . 
	\item  Training the classifiers based on 10 OCT volumes of each class. 
	\item Training the classifiers based on 14 OCT volumes of each class. 
\end{itemize}

In the last one which is also known as leave-three-out cross-validation, we randomly excluded one OCT volume of each class and use that as a test volume. The rest of the data including 14 volumes of each class were used for training the classifier. This process repeated until all 45 volumes were participated in the test phase.  Considering that discriminative signs of abnormalities do not appear in all B-scans of a volume, only 10 B-Scans of each volume which had a better representation of its class were inserted in the training phase . The process of two other experiments is similar to the above mentioned with different number of training data. The results of these 3 kinds of experiments are summarized in Tables \ref{tab01}-\ref{tab03}. As it was expected, increasing the number of training samples was resulted in improving the accuracy of classifiers. For the rest of the paper and final comparison, we focused on the leave-three-out cross-validation results of COPAR, FDDL and LRSDL classifiers. 

By considering the fact that dictionary learning methods use random initialization, after providing essentials for training phase, to avoid bias and examination of repeatability of the method, 15 repetitions was conducted. The following reported results are the mean results of 15 runs. 

It is worth mentioning that like other 3 dimensional imaging technique, considering only a single B-scan of an SD-OCT volume, is not sufficient for ascertaining retina diseases \cite{13}. For example, the existence of at least 50 distinct medium-sized drusen has introduced as a criterion for  diagnosis of intermediate dry-AMD. Since drusen can appear on several B-scans, it is more recommended that analysis of multi-frame should be attended in retina disease diagnosis. 

\begin{figure}[!t]
	\includegraphics[scale=.5]{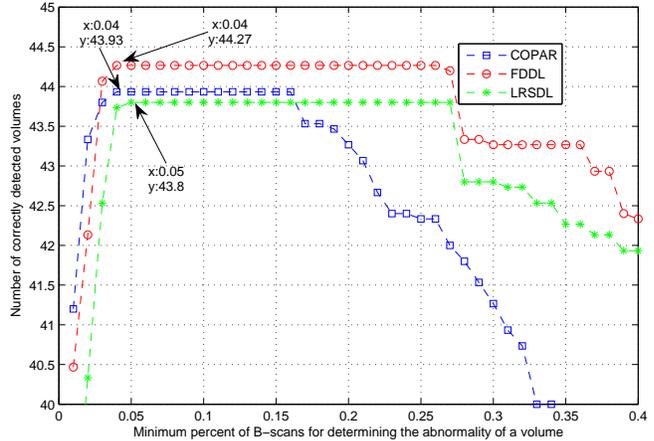}
	\caption{The result of SD-OCT volumes classification for different dictionary learning methods.\label{fig_5}}
\end{figure}

		
Another notable point is that although in some cases all B-scans of a volume, have signs of diseases, but the number of abnormal B-scans as the rate of abnormality detection should be as low as possible for cases at the very early stages of AMD and DME. So, determining a minimum number of B-scans consist of disease signs in an automatic diagnosis approach is challenging. In order to find the best criteria for our method, we examined different percent of abnormal B-scans for defining a volume as an abnormal case. The number of correct volumes tagged ( from the whole data set which consisted of 45 volumes) with respect to the minimum percent of B-scans which could be selected for data labelling has shown in Fig. \ref{fig_5}. In another word by increasing the minimum percent of B-scans for labelling a volume, from 0-40 percent and counting the correctly detected number of subjects, we found out that 4\% is the best criteria in our method for achieving the best result. Fig. \ref{fig_5} shows the results of these experiments for three different classifiers. As it was mentioned before, we repeated cross validation experiments 15 times, so the mean of the results have been shown in the Fig. \ref{fig_5}. Based on the calculated accuracy, reported in Table \ref{tab1} and Fig. \ref{fig_5}, we choose FDDL as the best classifier of our procedure. 

A comparison among the training time of our algorithm based on different classifiers also shows the privilege of FDDL. We performed our experiments, using MATLAB software on a desktop computer with Core i7-2630QM CPU at 2 GHz, and 4 GB of RAM. The computation time of leave-three-out cross-validation experiments based on COPAR, FDDL and LRSDL were 41, 22 and 32 minutes, respectively. Each leave-three-out cross-validation experiment includes 15 repetitions of training by 42 volumes and 3 volumes as test data. Most of the mentioned runtime devoted to training task and so the results show that the FDDL algorithm can learn dictionary faster than the two others. 

Three studies which evaluated the classification problem by individual subjects based on the same data set, are \cite{2},\cite{13, 14, 15}. Srinivasan, et al. in \cite{13} proposed a method which is based on multi-scale HOG feature extraction of OCT images and using the SVM algorithm as their classifier. Their reported results are based on performing 45 experiments using the leave-three-subjects-out cross validation strategy. Wang Yu et al. in \cite{14}, used the linear configuration pattern (LCP) based features of the OCT images which were followed by the Correlation-based Feature Subset (CFS) selection algorithm and chose their best model based on the sequential minimal optimization (SMO) algorithm. In order to classify SD-OCT images Yankui Sun et al. in \cite{15} suggest to partition every cropped image in the training set into small patches and then training a sparse dictionary with all SIFT descriptors extracted from the selected patches of the training images. Next, they obtain global representation for each OCT image by using an SP image representation, sparse coding, and max-pooling. Finally, by training three two-class linear SVMs, input images would be classified.  

Considering the similar preprocessing steps of \cite{13} and \cite{15}, we summarized their approaches in contrast to ours in Table \ref{tab0}.

In \cite{13}, the authors reported their best results based on considering 33\% or more of the B-scans of a volume as a criterion for classification of AMD/DME/normal data set. In \cite{14} and \cite{15}, a rule of the majority was utilized to determine the label of an SD-OCT volume. In other words, when most of the B-scans of a volume belonging to a specific class, that class is the final predicted label of the subject. Unfortunately, they did not mention exactly a minimum percent as their criterion for labeling a volume. 

Table \ref{tab2} shows the final classification accuracy of Srinivasan et al. \cite{13}, Wang Yu et al. \cite{14} and Yankui Sun et al. \cite{15} in contrast to our best results. It shows that while Srinivasan's method and Yankui Sun's method were not completely successful at identify normal cases from DME and AMD cases, our method could reach the accuracy of 100\% . In this field the performance of Wang Yu method is equal to ours. Furthermore, our method could thoroughly detect AMD cases like Yankui Sun and Srinivasan which is outperformed of Wang Yu. While three other methods were completely successful at identifying DME cases, our method did not succeed as others. Considering the whole dataset consist of normal, DME and AMD cases, the total performance of our proposed method was superior to three others by the accuracy of 98.37\%. 

\section{Conclusion}\label{sec4}

This study could provide a framework for the classification of two retina diseases against the normal subjects using the SD-OCT images. To achieve this goal we emphasized on machine learning methods with no acceptance of retina layer segmentation risk. In another word, considering the fact that retina layer segmentation is risky especially in the case of abnormal images, our proposed method is completely independent of retina layer segmentation. We propose to extract HOG feature of AMD, DME, and normal OCT images, due to the high capacity of this feature in illustrating appearing fold in retina layer in AMD and its ruptures in DME and flatness in normal B-Scans. 
In this study, we discussed 3 different kinds of dictionary learning methods for learning the highest discriminative atoms to classify AMD-DME-Normal images. As it was reported in the latter section, FDDL in conjunction with HOG feature is the best framework to achieve this goal, but it is worth mentioning that LRSD and COPAR method also could achieve the accuracy of more than 97.3\% which is comparable with previous studies. So, we emphasize that dictionary learning methods are powerful tools in discrimination of 3 investigated classes. 
Early diagnosis of diseases is another important issue that has been addressed in this study. We illustrated that using dictionary learning method have resulted in the diagnosis of abnormality of a subject based on abnormal detection of only 4\% of all B-scans of a volume. This percentage expresses the ability of the method to detect diseases in their early stages.


%
\bibliographystyle{imaiai}

\begin{table}[!b]
	\caption {Comparison of classification accuracy using COPAR classifier}\label{tab01}
	\begin{tabular}{lllll}
		& Training by 				& Training by 				&Training by\\ 
		Class								& 5 volumes  				    & 10 volumes 			    & 14 volumes\\ 
		& of each class 				& of each class 				& of each class \\
		\hline
		Normal							&	96.89\%					&97.33\%						&100\%\\
		DME								&	73.78\%				    &94.22\%						&92.86\%	\\			
		AMD								&	85.33\%					&95.11\%						&100\%\\
		\hline
		Whole data set  				&	85.33\%					&95.56\%						&\textbf{97.62\%}\\
	\end{tabular}
\end{table}		

\begin{table}[]
	\caption {Comparison of classification accuracy using FDDL classifier}\label{tab02}
	\begin{tabular}{lllll}
		& Training by 				& Training by 				&Training by\\ 
		Class								& 5 volumes  				    & 10 volumes 			    & 14 volumes\\ 
		& of each class 				& of each class 				& of each class \\
		\hline
		Normal							&93.78\%							&100\%						&100\%\\
		DME								&92.00\%						    &95.56\%	  					&95.13\%	\\
		AMD								&98.22\%	  						&99.56\%						&100\%\\
		\hline
		Whole data set  				&	94.67\%					    &\textbf{98.37\%}\%						    &\textbf{98.37\%}	\\
	\end{tabular}
\end{table}	

\begin{table}[]
	\caption {Comparison of classification accuracy using LRSDL classifier}\label{tab03}
	\begin{tabular}{lllll}
		& Training by 				& Training by 				&Training by\\ 
		Class								& 5 volumes  				    & 10 volumes 			    & 14 volumes\\ 
		& of each class 				& of each class 				& of each class \\
		\hline
		Normal							&85.33\%							&97.78\%						&100\% \\
		DME	    						&93.33\%						    &92.89\%	 					&92.00\%	  \\
		AMD								&97.78\%							&99.11\%						&100\% \\
		\hline 
		Whole data set  				&92.15\%							&96.59\%						&\textbf{97.33\%}		\\
	\end{tabular}
\end{table}	
\begin{landscape}
\begin{table}[]
	\caption{The results of different dictionary learning classifiers  based on leave-tree-out cross validation experiments.}\label{tab1}
	\begin{tabular}{lllll}
		Method     &Normal  &DME                     &AMD		   &Whole data set \\
		\hline
		COPAR      &100\%   &92.86\%                 & 100\% 		& 97.62\% \\
		FDDL	   &100\%	&\textbf{95.13\%}        &100\% 		&  \textbf{98.37}\%\\
		LRSDL      &100\%   &92\%	                 &100\%         &  97.33\%\\
	\end{tabular}
\end{table}

\begin{table}[]
	\caption{Comparison of classification accuracy of four studies.}\label{tab2}
	\begin{tabular}{lllll}
		&Srinivasan                 & Wang Yu               & Yankui Sun          & This study\\
		&et al. \cite{13}           & et al. \cite{14}      &  et al. \cite{15}    &                \\
		\hline
		Normal &	86.66\%&	100\%	&93.33\%  &100\%\\	
		DME	&100\%	&100\%	&100\%	&95.13\%\\
		AMD	&100\%	&93.33\%&	100\%	&100\%\\
		\hline
		Whole data set	&95.55\%	&97.78\%	&97.78\%	& \textbf{98.37\%}\\		
	\end{tabular}
\end{table}

	\begin{table}
	\caption {The results of different dictionary learning classifiers  based on leave-tree-out cross validation experiments.}\label{tab0}
	\begin{tabular}{llll}
			Classification steps 	& Srinivasan et al.\cite{13}                &Yankui Sun  et al. \cite{15}	      &This study\\
			\hline
			Denoising 			&Using Block-matching and                   &Using BM3D	                           &The proposed method of \cite{16_1}\\
								&	3D filtering	(BM3D) 				    &									   &  \\
			\hline							   
			Flatten retinal &Aligned retina regions by fitting a     & Extracting two sets of data points of image		& Aligned retina regions by fitting \\
			curvature 	    & $2^{nd}$ order polynomial to the RPE layer& and automatically choosing one set of  	& a $1^{st}$ to $4^{th}$ order polynomial to   \\ 
			& and then flattened the retina.	      & points which is more representative of  	        & the RPE layer points and then \\	
			&			                                          & retina	and then choosing a $2^{nd}$ order   &           flattened the retina.  \\
			&			                                            & polynomial or straight line to fit the points.     &                       \\
						\hline							
			Crop images 	            &45*150	                                        &The size of cropped image was not      &  65*380\\
			&														&mentioned exactly.							  &\\
						\hline 
			Feature extraction 	    &Histogram of Oriented Gradients               & Spatial pyramid matching based 			&HOG\\
			& (HOG)  			                                       & on sparse codes of scale-invariant			& \\
			&                                                              &feature transform			                        & \\
						\hline 						                                                                       
			Classifier                    &Support Vector Machine (SVM)	        &SVM							&Fisher Discrimination Dictionary\\
			&														&                                  &Learning (FDDL)\\
	\end{tabular}
	\end{table}
\end{landscape}

\end{document}